\documentclass[11pt,twoside]{article}

\usepackage{graphicx}
\usepackage{asp2006}
\usepackage{epsf}
\usepackage{lscape}

\begin{document}
\title{Galaxies and Groups in the Local Supercluster}

\author{D. Makarov and I. Karachentsev}
\affil{Special Astrophysical Observatory  Russian Academy Sciences,
Nizhnij Arkhyz, Russia}

\section{ALGORITHM}
Our clusterization algorithm bases on assumption that a total energy
of physical pair of galaxies must be negative. Each pair of galaxies
is examined to obey the conditions:
$$
\frac{V^2_{12}R^3_{\perp}}{2GM_{12}}<1,\hspace{0.5cm}
\frac{\pi H^2R^3_{\perp}}{8GM_{12}}<1
$$
First equation represents the kinetic and gravitational
energy ratio, where $V_{12}$ is velocity difference, $R_{\perp}$ is
projected linear separation and $M_{12}$ is total mass of pair.
Second equation limits maximal linear separation of galaxies in
physical pair. As natural bound we use zero-velocity surface
which separates the collapsing volume against expanding space (Sandage, 2007).

We determine the masses of galaxies from their $K$-band luminosity
assuming that all galaxies have the same mass-to-luminosity ratio
$M=6L_K$.
If galaxies in the pair have a common component, we use the criteria
above to combine the resulting pairs into triplets, quartets and so
on. We repeat these steps until there are objects to combine into
groups.

\section{SAMPLE}
Our main sources of the data on redshifts, apparent magnitudes,
morphological types, and other parameters of galaxies are the
HyperLEDA and NED databases. We collected the 10915 galaxies with
velocity $<3500$ km/s respect to the Local Group and located at Galactic
latitudes $\mid b\mid <15\deg$. Many galaxies were inspected visually to reject
objects with wrong velocities or false multiplicity. This astronomical
spam consists significant part (at least 10\%) of real galaxies in
considered volume. The sample includes the entire Local Supercluster
(with a center at $V_{LG}\sim$1200 km/s) together with its distant outskirts,
neighboring voids, and spurs of neighboring clusters.

We determined the blue magnitude $B_T$  and type $T$ of galaxies with
unknown parameters by visual comparison of these objects with
reference galaxies of similar structure found on the digital version
of POSS. We converted the estimates of galaxy magnitudes in optical
($V, R, I$) and near-infrared bands ($J, H$) into the $K$-band magnitude
using synthetic galaxy colors of Bizzoni (2005) and Fukujita et al. (1995). 
We use relation between $B-K$ color and morphological type discussed by 
Jarett et al. (2003) and Karachentsev \& Kut'kin (2005).

\section{GROUPS}
The 5927 galaxies of 10915 (about 54\%) have been gathered in 1082 groups. 
The 46\% of galaxies of our sample are single objects. 
The 9\% (1032 galaxies) form a pairs. 
The triplets comprise about 5\% (513 objects) of all galaxies.

\begin{figure}[!ht]
\begin{center}
\includegraphics[scale=0.35]{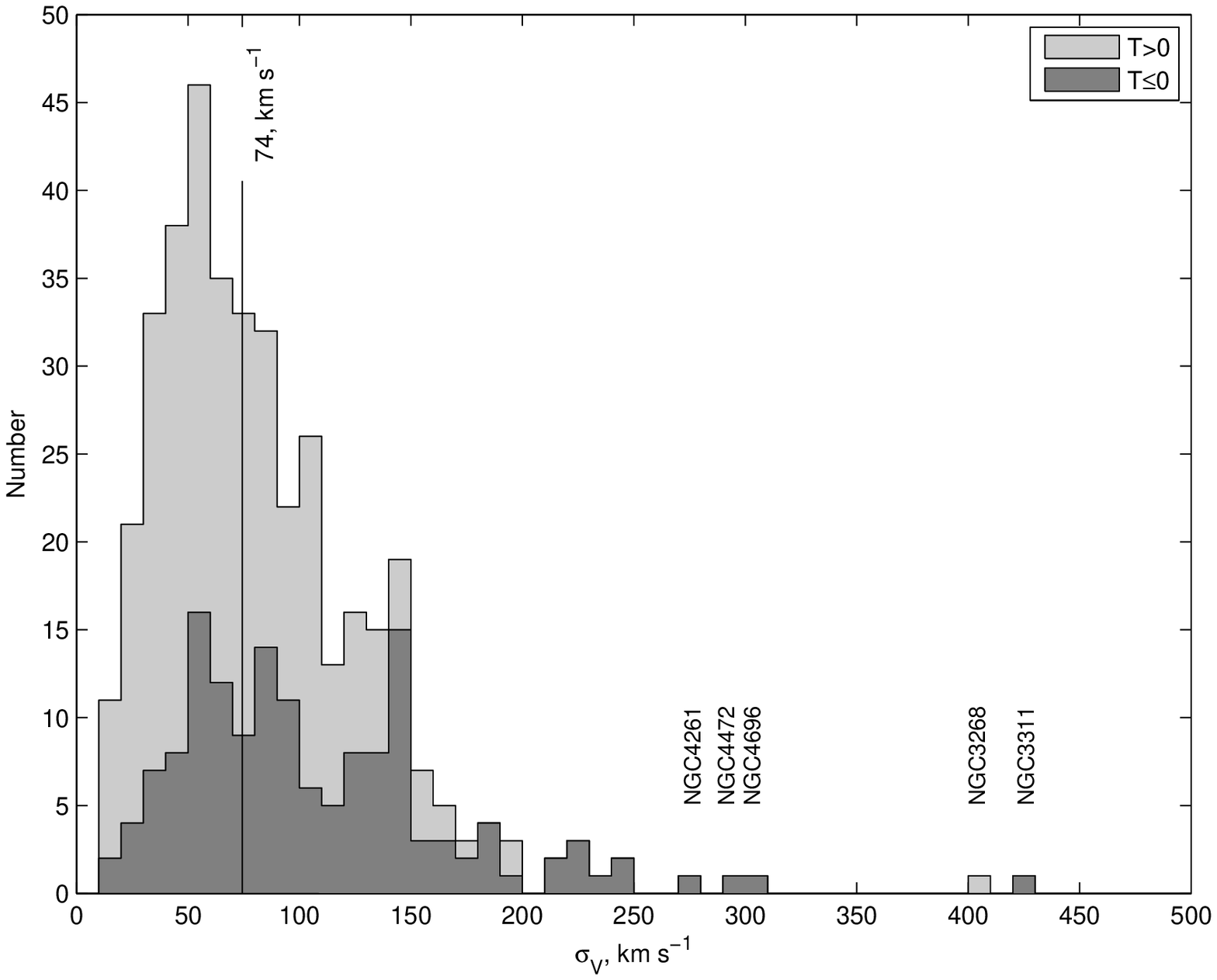}
\includegraphics[scale=0.35]{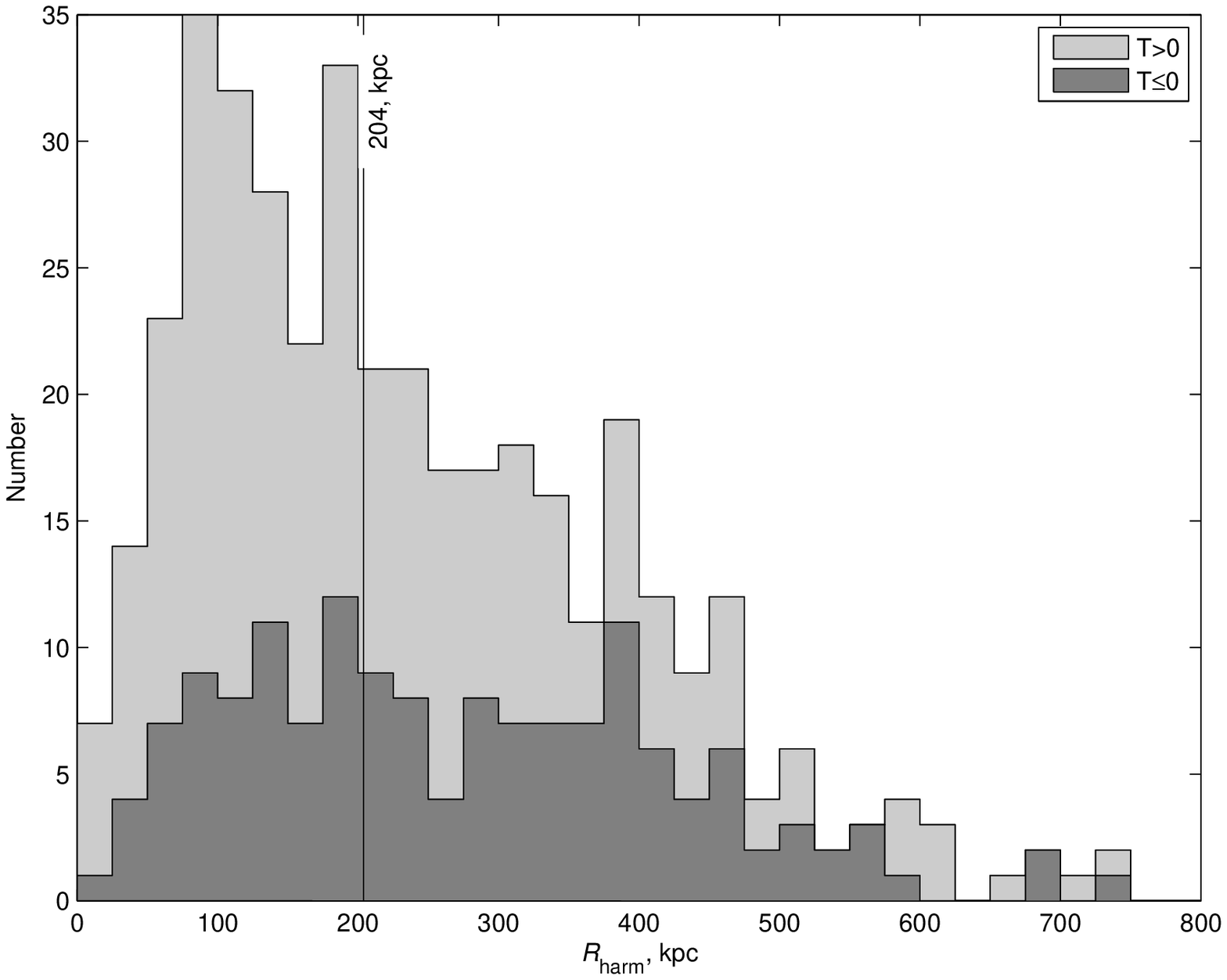}
\end{center}
\caption{The distribution of the groups by dispersion (left panel) and harmonic radius (right panel).}
\end{figure}

The well populated groups with number of members bigger than 4 in the Local Supercluster sample are 
characterized by velocity dispersion 74 km/s and harmonic radius 204 kpc. 
These values are typical for small groups like Local Group of galaxies, M81 and Centaurus A.

\begin{figure}[!ht]
\begin{center}
\includegraphics[scale=0.5]{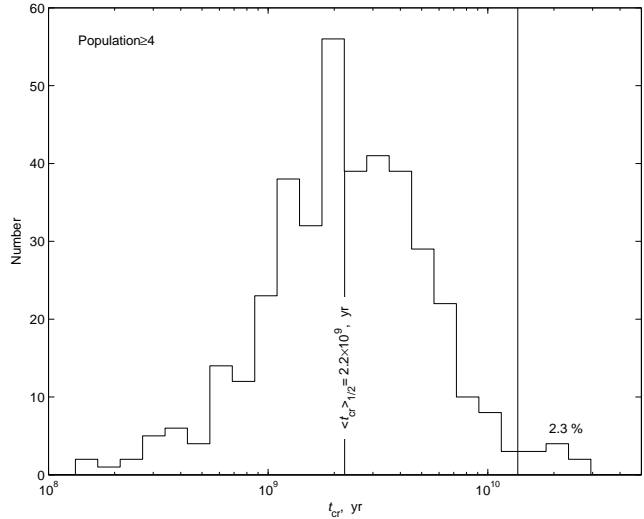}
\end{center}
\caption{The distribution of the groups by crossing time.}
\end{figure}

\begin{figure}[!ht]
\begin{center}
\includegraphics[scale=0.6]{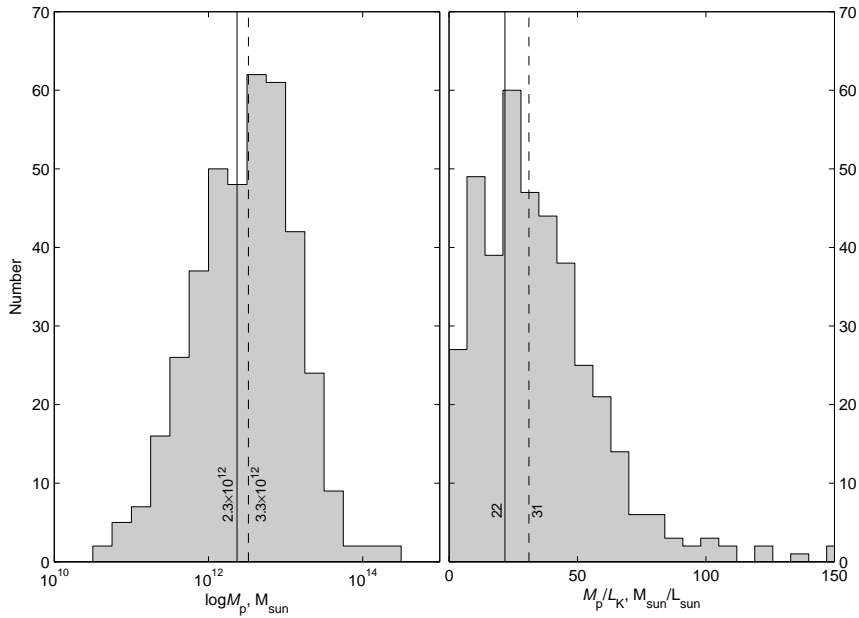}
\end{center}
\caption{The distribution of the groups by total mass (left panel) and mass-to-light ration (right panel).}
\end{figure}

The crossing time of most of groups is less than the age of the Universe. 
It indicates that groups under consideration are in a virialized state. 
We found the median corrected mass-to-$K$-luminosity ratio for groups to be 22 in solar units. 
This values shows presence of moderate amount of dark matter in considered groups. 
It is significantly higher then mass-to-light ratio for individual galaxies,
this value is comparable to $M/L_K$ ratio of the small groups like nearby
groups in the Local Volume and it is about half of ratio for clusters (Karachentsev \& Kut'kin, 2005).
The mass-to-luminosity ratio of groups in Local Supercluster is significantly lower than mean
global value of $M/L_K\sim90$ which is expected in standard cosmology with
$\Omega_m=0.27$ and $\Omega_{\Lambda}=0.73$ (Spergel, 2007).

\acknowledgements This work has been supported by Russian-Ukrainien
grant 09--02--90414, RFBR grants: 07--02--00005,  07--02--00792, 08--02--00627.

{}

\end{document}